\input{epsf}

\documentstyle[fortschritte]{article}
\input{tcilatex}
\input tcilatex

\begin{document}

\author{D. Vion, A. Aassime, A. Cottet, P. Joyez, H. Pothier, \and C.
Urbina, D. Esteve and M.H. Devoret\thanks{%
Present address: Applied Physics Department, Yale University, New Haven, CT
06520, USA} \\
%EndAName
Quantronics Group, Service de Physique de l'Etat Condens\'{e},\\
D\'{e}partement des Sciences de la Mati\`{e}re, CEA-Saclay\\
91191 Gif-sur-Yvette, France}
\title{Rabi oscillations, Ramsey fringes and spin echoes in an electrical
circuit}
\date{}
\maketitle

\begin{abstract}
We present a superconducting tunnel junction circuit which behaves as a
controllable atom, and whose ground and first excited state form an
effective spin 1/2. By applying microwave pulses, we have performed on this
circuit experiments demonstrating the controlled manipulation of the spin :
Rabi precession, Ramsey interferences, and spin echoes.
\end{abstract}

\bigskip The state variables of an electrical circuit, like voltages and
currents can be made to behave quantum mechanically by minimizing the
coupling to external degrees of freedom through a proper design. Circuits
based on superconducting tunnel junctions have displayed signatures of
macroscopic quantum behavior \cite%
{MQT,Bouchiat98,Nakamura,Mooij,Lukens,Han,Martinis}, but the level of
coherence of the quantum states remained until now much smaller than for
isolated atoms or ions. In the ``quantronium'' circuit presented here, a
coherence quality factor of more than $10^{4}$ has been obtained, thus
allowing the coherent manipulation of the state of the system like in atomic
physics and NMR experiments \cite{science}.

The quantronium consists of a superconducting loop interrupted by two
adjacent small Josephson tunnel junctions with capacitance $C_{j}/2$ and
Josephson energy $E_{J}/2$ each, which define a low capacitance
superconducting electrode called the ``island'', and by a large Josephson
junction with large Josephson energy $(E_{J0}\approx 20E_{J})$ (see Fig. 1).
The island is biased by a voltage source $U$ through a gate capacitance $%
C_{g}$. In addition to $E_{J}$, the quantronium has a second energy scale
which is the Cooper pair Coulomb energy $E_{CP}=(2e)^{2}/2\left(
C_{g}+C_{j}\right) $. The temperature $T$ and the superconducting gap $%
\Delta $ satisfy $k_{B}T\ll \Delta /\ln {\cal N}$ and $E_{CP}<\Delta
-k_{B}T\ln {\cal N}$, where ${\cal N}$ is the total number of paired
electrons in the island. The number of excess electrons is then even \cite%
{Tuominen,parity}, and the system has discrete quantum states which are in
general quantum superpositions of several charge states with different
number $\hat{N}$ of excess Cooper pairs in the island. Neglecting the loop
inductance and the charging energy of the large junction, the Hamiltonian of
the circuit is

\begin{equation}
\hat{H}=E_{CP}\left( \hat{N}-N_{g}\right) ^{2}-E_{J}\cos (\frac{\hat{\gamma}%
+\phi }{2})\cos \hat{\theta}-E_{J0}\cos \hat{\gamma}-I_{b\ }\varphi _{0}\ 
\hat{\gamma}\ \text{,}  \label{Ham}
\end{equation}%
where $N_{g}=C_{g}U/2e$ is the dimensionless gate charge, $\phi =\Phi
/\varphi _{0}$ is a phase bias, with $\Phi $ the external flux imposed
through the loop and $\varphi _{0}=\hbar /2e$, $\hat{\gamma}$ is the phase\
across the large junction, and $\hat{\theta}$ is the phase operator
conjugate to the Cooper pair number $\hat{N}$. The bias current $I_{b}$ is
zero except during readout of the state \cite{quantro}.

In our experiment $E_{J}\simeq E_{CP}$ and neither $\hat{N}$ nor $\hat{\theta%
}$ is a good quantum number. In contrast, the large junction is shunted by a
large capacitor $C$ so that $\hat{\gamma}$ is almost a classical variable.
In this regime, the energy spectrum is sufficiently anharmonic for the two
lowest energy states $\left| 0\right\rangle $ and $\left| 1\right\rangle $
to form a two-level system. This system corresponds to an effective spin
one-half $\vec{s}$ with eigenstates $\left| 0\right\rangle \,\equiv \left|
s_{z}=+1/2\right\rangle $ and $\left| 1\right\rangle \equiv \left|
s_{z}=-1/2\right\rangle .$ At $N_{g}=1/2,$ $I_{b}=0$ and $\phi =0,$ its
``Zeeman energy'' $h\nu _{01}$, of the order of{\em \ }$E_{J},$ is
stationary with respect to $N_{g}$, $I_{b}$ and $\phi $ \cite{science},
making the system immune to first order fluctuations of the control
parameters. Manipulation of the quantum state is thus performed at this
optimal point by applying microwave pulses $u(t)$ with frequency $\nu \simeq
\nu _{01}$ to the gate, and any superposition $\left| \Psi \right\rangle
=\alpha \left| 0\right\rangle +\beta \left| 1\right\rangle $ can be
prepared, starting from $\left| 0\right\rangle $. In a frame rotating around
the quantization axis $z$ at frequency $\nu ,$ the microwave voltage acts on 
$\vec{s}$ as an effective dc magnetic field in the $x~z$ plane, with $x$ and 
$z$ components proportional to the microwave amplitude and to the detuning $%
\Delta \nu =\nu -\nu _{01},$ respectively.

\begin{figure}
\centerline{\epsfxsize=5.815in \epsfysize=2.5529in
\epsffile{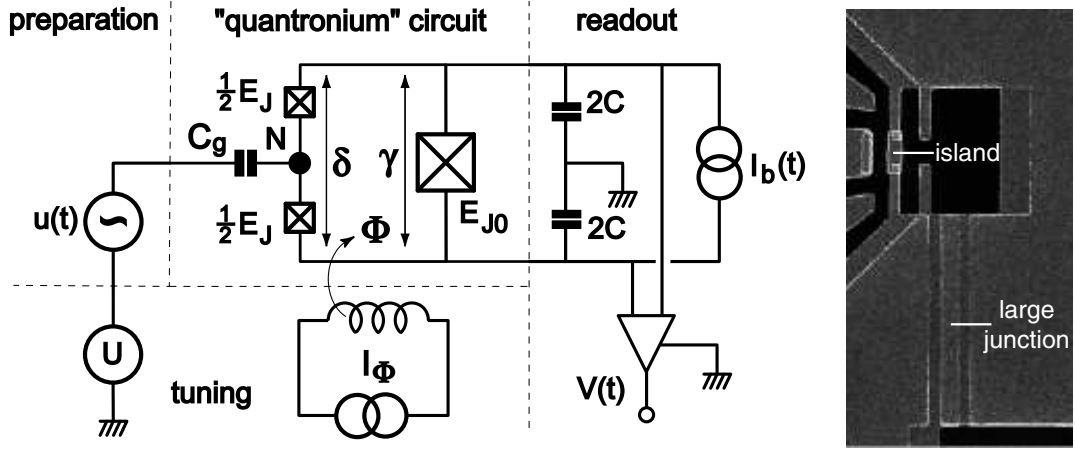} }
\caption{Left : Idealized circuit diagram of the ``quantronium'',
a quantum coherent circuit with its tuning, preparation and readout blocks.
The circuit consists of an island (black node) delimited by two small
Josephson junctions (crossed boxes) in a superconducting loop. The loop also
includes a third, much larger Josephson junction shunted by a capacitance $C$%
. The Josephson energies of the island and of the large junction are $E_{J}$
and $E_{J0}$. The Cooper pair number in the island $N$ and the phases $%
\protect\delta $ and $\protect\gamma $ are the degrees of freedom of the
circuit. A dc voltage $U$ applied to the gate capacitance $C_{g}$ and a dc
current $I_{\protect\phi }$ applied to a coil producing a flux $\Phi $ in
the circuit loop tune the quantum energy levels. Microwave pulses $u(t)$
applied to the gate prepare arbitrary quantum states of the circuit. The
states are readout by applying a current pulse $I_{b}(t)$ to the large
junction and by monitoring the voltage $V(t)$ across it. Right: Scanning
electron micrograph of a sample.}
\end{figure}

%\FRAME{dhFUw}{5.815in}{2.5529in}{%
%0pt}{\Qcb{Figure 1: Left : Idealized circuit diagram of the ``quantronium'',
%a quantum coherent circuit with its tuning, preparation and readout blocks.
%The circuit consists of an island (black node) delimited by two small
%Josephson junctions (crossed boxes) in a superconducting loop. The loop also
%includes a third, much larger Josephson junction shunted by a capacitance $C$%
%. The Josephson energies of the island and of the large junction are $E_{J}$
%and $E_{J0}$. The Cooper pair number in the island $N$ and the phases $%
%\protect\delta $ and $\protect\gamma $ are the degrees of freedom of the
%circuit. A dc voltage $U$ applied to the gate capacitance $C_{g}$ and a dc
%current $I_{\protect\phi }$ applied to a coil producing a flux $\Phi $ in
%the circuit loop tune the quantum energy levels. Microwave pulses $u(t)$
%applied to the gate prepare arbitrary quantum states of the circuit. The
%states are readout by applying a current pulse $I_{b}(t)$ to the large
%junction and by monitoring the voltage $V(t)$ across it. Right : Scanning
%electron micrograph of a sample.}}{}{setup.eps}{\special{language
%"Scientific Word";type "GRAPHIC";maintain-aspect-ratio TRUE;display
%"USEDEF";valid_file "F";width 5.815in;height 2.5529in;depth
%0pt;original-width 7.0802in;original-height 7.3189in;cropleft "0";croptop
%"1";cropright "1";cropbottom "0";filename 'setup.eps';file-properties
%"XNPEU";}}

For readout, we have implemented a strategy reminiscent of the Stern and
Gerlach experiment \cite{Stern}, in which the information about the spin of
silver atoms is transferred onto their transverse position$.$ In our
experiment, the information on $\vec{s}$ is transferred onto the phase $\hat{%
\gamma},$ and the two states are discriminated through the supercurrent in
the loop $\left\langle \hat{I}\right\rangle =\left\langle \partial \hat{H}%
/\partial \hat{\delta}\right\rangle /\varphi _{0},$ where{\em \ }$\hat{\delta%
}=\hat{\gamma}+\phi $ is the phase difference across the series combination
of the two small junctions \cite{design,Zorin,Buisson}. For this purpose, a
trapezoidal readout pulse $I_{b}(t)$ with a peak value slightly below the
critical current $I_{0}=E_{J0}/\varphi _{0}$ is applied to the circuit (see
Fig. 2).

\begin{figure}
\centerline{\epsfxsize=2.8176in \epsfysize=1.8775in
\epsffile{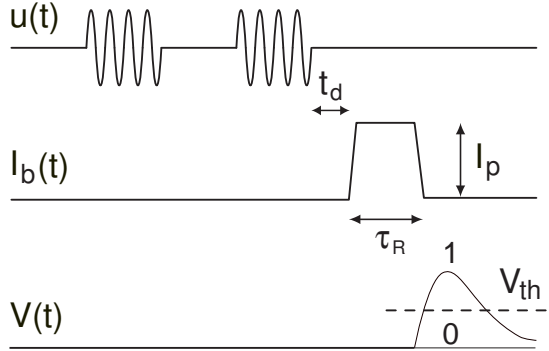} }
\caption{Signals
involved in quantum state manipulations and measurement of the
``quantronium''. Top: microwave voltage pulses are applied to the gate for
state manipulation. Middle: a readout current pulse $I_{b}(t)$ with
amplitude $I_{p}$ is applied to the large junction $t_{d}$ after the last
microwave pulse. Bottom: voltage $V(t)$ across the junction. The occurence
of a pulse depends on the occupation probabilities of the energy
eigenstates. A discriminator with threshold $V_{th}$ converts $V(t)$ into a
boolean 0/1 output for statistical analysis.}
\end{figure}

%\FRAME{dhFUw}{}{}{0pt}{\Qcb{Figure 2: }}{}{}{\special%
%{language "Scientific Word";type "GRAPHIC";maintain-aspect-ratio
%TRUE;display "USEDEF";valid_file "F";width 2.8176in;height 1.8775in;depth
%0pt;original-width 3.9851in;original-height 2.629in;cropleft "0";croptop
%"1";cropright "1";cropbottom "0";filename 'signal.eps';file-properties
%"XNPEU";}}

When starting from $\langle \hat{\delta}\rangle \approx 0$, the
phases $\langle \hat{\gamma}\rangle $ and $\langle \hat{\delta}\rangle $
grow during the current pulse, and consequently an $\vec{s}$-dependent
supercurrent develops in the loop. The loop current is the analog of the
transverse acceleration experienced by a silver atom in the magnetic field
gradient. This current adds to the bias-current in the large junction, and
by precisely adjusting the amplitude and duration of the $I_{b}(t)$\ pulse,
the large junction switches during the pulse to a finite{\em \ }voltage
state with a large probability $p_{1}$ for state $\left| 1\right\rangle $
and with a small probability $p_{0}$ for state $\left| 0\right\rangle $%
\thinspace \cite{design}. A switching event corresponds to the impact of a
silver atom on the top spot of the screen. The absence of switching
corresponds to the impact of a silver atom on the bottom spot of the screen.
For the parameters of the experiment, the efficiency of this projective
measurement should be $\eta =p_{1}-p_{0}=0.95$ for optimum readout
conditions. Large ratios $E_{J0}/E_{J}$ and $C/C_{j}$ provide further
protection from the environment.

An actual ``quantronium'' sample is shown on the right side of Fig.~1. It
was fabricated with standard technique of aluminum evaporation through a
shadow-mask obtained by e-beam lithography. With an external microwave
capacitor $C=1~{\rm pF}$, the plasma frequency of the large junction with $%
I_{0}=0.77~{\rm 
%TCIMACRO{\U{b5}}%
%BeginExpansion
{\mu}%
%EndExpansion
A}$ is $\omega _{p}/2\pi \simeq 8~{\rm GHz}$. The sample and last filtering
stage were anchored to the mixing chamber of a dilution refrigerator with 15
mK base temperature. The switching of the large junction to the voltage
state is detected by measuring the voltage across it with a room temperature
preamplifier followed by a discriminator with a threshold voltage $V_{th}$
well above the noise level (Fig. 2). By repeating the experiment (typically
a few 10$^{4}$ times), we can determine the switching probability, and
hence, the occupation probabilities $\left| \alpha \right| ^{2}$ and $\left|
\beta \right| ^{2}$.

The readout part of the circuit was tested by measuring the switching
probability $p$ as a function of the pulse height $I_{p}$, for a current
pulse duration of $\tau _{r}=100~{\rm ns}$, at thermal equilibrium . The
discrimination between the currents corresponding to the $\left|
0\right\rangle $ and $\left| 1\right\rangle $ states was found to have an
efficiency of $\eta =0.6$, which is lower than the expected $\eta =0.95$.
Measurements of the switching probability as a function of temperature and
repetition rate indicate that the discrepancy between the theoretical and
experimental readout efficiency could be due to an incomplete thermalization
of our last filtering stage in the bias current line.

\begin{figure}
\centerline{\epsfxsize=5.9395in \epsfysize=2.8409in
\epsffile{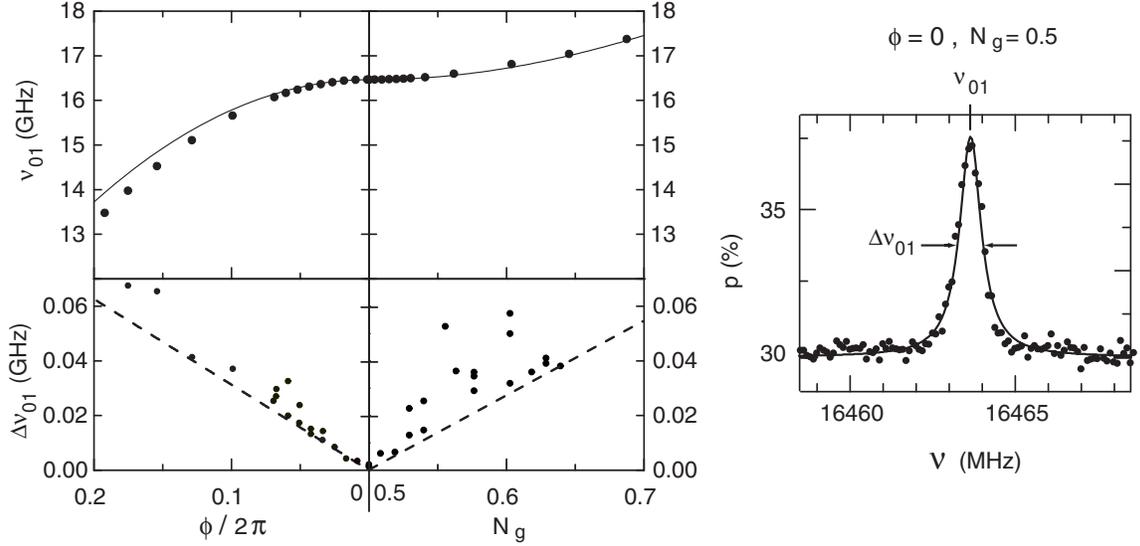} }
\caption{Left: Measured center
frequency $\protect\nu _{01}$ (top panels, symbols) and full width at half
maximum $\Delta \protect\nu _{01}$ (bottom panels, symbols) of the resonance
as a function of reduced gate charge $N_{g}$ for reduced flux $\protect\phi %
=0$ (right panels), and as a function of $\protect\phi $ for $N_{g}=0.5$
(left panels), at $15{\rm ~mK}$. Spectroscopy is performed by measuring the
switching probability $p$ when a continuous microwave irradiation of
variable frequency is applied to the gate before readout ($t_{d}<100~{\rm ns}
$). Continuous lines in top panels: theoretical best fit (see text). Dotted
lines in bottom panels correspond to $\partial \Delta \protect\nu %
_{01}/\partial N_{g}\simeq 250~{\rm MHz}$ and $\partial \Delta \protect\nu %
_{01}/\partial (\protect\phi /2\protect\pi )\simeq 430~{\rm MHz.}$ They give
a lower bound to the measured $\Delta \protect\nu _{01}.$ Right: Lineshape
measured at the optimal working point $\protect\phi =0$ and $N_{g}=0.5$
(dots). Lorentzian fit with a FWHM $\Delta \protect\nu _{01}=0.8~{\rm MHz}$
and a center frequency $\protect\nu _{01}=16463.5~{\rm MHz}$ (solid line).}
\end{figure}

%\FRAME{dhFUw}{}{}{0pt}{\Qcb{Figure 3: }}{%
%}{}{\special{language "Scientific Word";type
%"GRAPHIC";maintain-aspect-ratio TRUE;display "USEDEF";valid_file "F";width
%5.9395in;height 2.8409in;depth 0pt;original-width 10.562in;original-height
%7.9122in;cropleft "0";croptop "1";cropright "1";cropbottom "0";filename
%'spectro.eps';file-properties "XNPEU";}}

Spectroscopic measurements of $\nu_{01}$ were performed 
by applying to the gate a weak continuous microwave
irradiation suppressed just before the readout current pulse. The variations
of the switching probability as a function of the irradiation frequency
display a resonance whose center frequency evolves as a function of the dc
gate voltage and flux as the Hamiltonian (1) predicts, reaching $\nu
_{01}\simeq 16.5~{\rm GHz}$ at the optimal working point (see Fig.~3). The
small discrepancy between theoretical and experimental values of the
transition frequency at nonzero magnetic flux is attributed to flux
penetration in the small junctions not taken into account in the model. We
have used these spectroscopic data to precisely determine the relevant
circuit parameters and found $i_{0}=18.1~{\rm nA}$ and $E_{J}/E_{CP}=1.27$.
The linewidth $\Delta \nu _{01}$ is given in the bottom panels in Fig. 3 as
a function of $\phi $ and $N_{g}$. At the optimal working point, the
linewidth was found to be minimal with a $0.8~$MHz full width at
half-maximum, corresponding to a quality factor $Q=2\times 10^{4}$. The
lineshape was found to be irreproducible, probably because of slight shifts
of the resonance frequency during the measurement, related to low frequency
charge or noise on the phase $\delta $. If one considers the narrowest lines
recorded, the linewidth varies linearly when departing from the optimal
point $(N_{g}=1/2,~\phi =0,~I_{b}=0)$, the proportionality coefficients
being $\partial \Delta \nu _{01}/\partial N_{g}\simeq 250~{\rm MHz}$ and $%
\partial \Delta \nu _{01}/\partial (\phi /2\pi )\simeq 430~{\rm MHz}$. These
values can be translated into RMS deviations $\Delta N_{g}=0.004$ and $%
\Delta (\delta /2\pi )=0.002$ during the time needed to record the
resonance. The residual linewidth at the optimal working point can be
explained by the second order contribution of these noises. The amplitude of
the charge noise is in agreement with measurements of $1/f$ charge noise %
\cite{PTB}, and its effect could be minimized by increasing the $E_{J}/E_{C}$
ratio. By contrast, the amplitude of the phase noise corresponds to a large
flux noise \cite{Wellstood}, but it could be also attributed to bias current
noise.

When varying the delay between the end of a resonant irradiation and the
measurement pulse at the optimal working point, the switching probability
decays with a time constant $T_{1}=1.8~{\rm \mu s}$ (see Fig. 4). Supposing
that the energy relaxation of the system is only due to the bias circuitry,
a calculation along the lines of Ref.~\cite{relax} predicts that $T_{1}\sim
10~{\rm 
%TCIMACRO{\U{b5}}%
%BeginExpansion
{\mu}%
%EndExpansion
s}$ for a crude discrete element model. This result shows that no
detrimental sources of dissipation have been seriously overlooked in our
circuit design.

\begin{figure}
\centerline{\epsfxsize=3.4022in \epsfysize=2.4967in
\epsffile{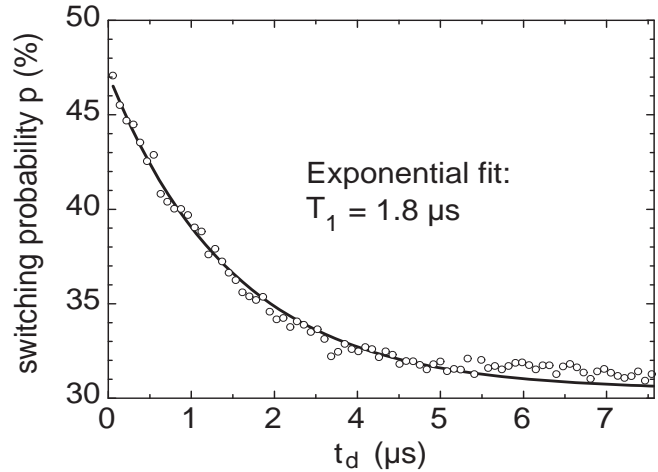} }
\caption{Decay
of the switching probability as a function of the delay time $t_{d}$ after a
continuous excitation at the center frequency of the resonance line. The
solid line is an exponential fit (vertically offset by the signal measured
without microwave applied to the gate), from which the relaxation time $%
T_{1}=1.8$~%
%TCIMACRO{\U{b5}}%
%BeginExpansion
$\mu$%
%EndExpansion
s is obtained.}
\end{figure}

%\FRAME{dhFUw}{}{}{0pt}{\Qcb{Figure 4: }}{}{}{\special{language "Scientific Word";type
%"GRAPHIC";maintain-aspect-ratio TRUE;display "USEDEF";valid_file "F";width
%3.4022in;height 2.4967in;depth 0pt;original-width 6.6521in;original-height
%4.8248in;cropleft "0";croptop "1";cropright "1";cropbottom "0";filename
%'t1.eps';file-properties "XNPEU";}}

We have then performed controlled rotations of $\vec{s}$ with large
amplitude microwave pulses. Prior to readout, a single pulse at the
transition frequency with variable amplitude $U_{%
%TCIMACRO{\U{b5}}%
%BeginExpansion
{\mu}%
%EndExpansion
w}$ and duration $\tau $ was applied. The resulting change in switching
probability is an oscillatory function of the product $U_{%
%TCIMACRO{\U{b5}}%
%BeginExpansion
{\mu}%
%EndExpansion
w}\tau $ (see Fig.~5), in agreement with the theory of Rabi oscillations %
\cite{Rabi}. It provides direct evidence that the resonance indeed
corresponds to an effective spin rather than to a spurious harmonic
oscillator resonance in the circuit. The proportionality ratio between the
Rabi period and $U_{%
%TCIMACRO{\U{b5}}%
%BeginExpansion
{\mu}%
%EndExpansion
w}\tau $ was used to calibrate microwave pulses for the application of
controlled rotations of $\vec{s}$.

\begin{figure}
\centerline{\epsfxsize=4.2523in \epsfysize=3.2145in
\epsffile{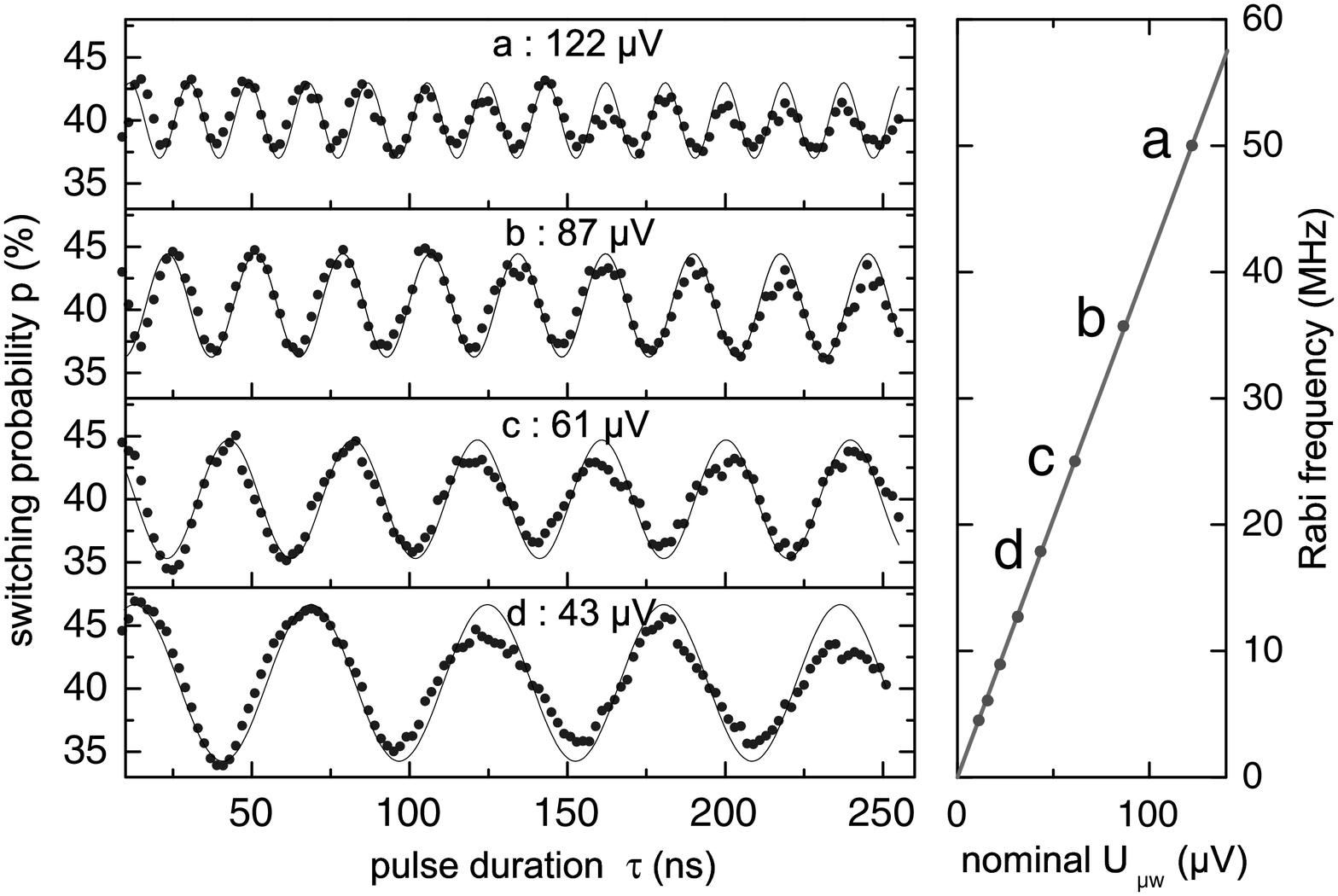} }
\caption{Left: Rabi oscillations of the switching probability $p$ ($%
5\times 10^{4}$ events) measured just after a resonant microwave pulse of
duration $\protect\tau $. Data taken at 15 mK for nominal pulse amplitudes $%
U_{%
%TCIMACRO{\U{b5}}%
%BeginExpansion
{\mu}%
%EndExpansion
w}=122{\rm ,}$ $87${\rm , }$61$, and $43~{\rm 
%TCIMACRO{\U{b5}}%
%BeginExpansion
{\mu}%
%EndExpansion
V}$ (dots, from top to bottom). The Rabi frequency is extracted from
sinusoidal fits (continuous lines). Right: the Rabi frequency (dots) varies
linearly with $U_{\protect\mu w}$, as expected. The four points labelled $a$
to $d$ correspond to the curves on the left.}
\end{figure}

%\FRAME{dhFUw}{4.2523in}{3.2145in}{0pt}{%
%\Qcb{Figure 5: }}{}{rabi.eps}{\special{language
%"Scientific Word";type "GRAPHIC";maintain-aspect-ratio TRUE;display
%"USEDEF";valid_file "F";width 4.2523in;height 3.2145in;depth
%0pt;original-width 297cm;original-height 210cm;cropleft "0";croptop
%"1";cropright "1";cropbottom "0";filename 'rabi.eps';file-properties
%"XNPEU";}}

The measurement of the coherence time of $\vec{s}$ during free evolution was
obtained by performing a Ramsey-fringes-like experiment \cite{Ramsey}. One
applies on the gate two phase coherent microwave pulses corresponding each
to a $\pi /2$ rotation around $x$ \cite{noteangle} and separated by a delay $%
\Delta t$ during which the spin precesses freely around $z$. For a given
detuning $\Delta \nu $ of the microwave frequency, the switching probability
displays decaying oscillations of frequency $\Delta \nu $ (see Fig.~6),
which correspond to the ``beating'' of the spin precession with the external
microwave field. The envelope of the oscillations yields the coherence time $%
T_{\varphi }\simeq 0.5~{\rm 
%TCIMACRO{\U{b5}}%
%BeginExpansion
{\mu}%
%EndExpansion
s}$. Given the transition period $1/\nu _{01}\simeq 60~{\rm ps}$, this means
that $\vec{s}$ can perform on average 8000 coherent free precession turns.%

\begin{figure}
\centerline{\epsfxsize=3.25in \epsfysize=2.5391in
\epsffile{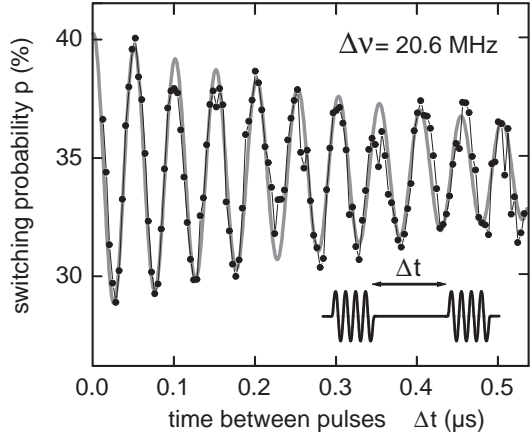} }
\caption{Ramsey fringes of the
switching probability $p$ after two $\protect\pi /2$ microwave pulses
separated by $\Delta t$. Dots: data at 15 mK. The total acquisition time was
5~mn. Continuous grey line: fit by exponentially damped sinusoid with
frequency $20.6~{\rm MHz}$, equal to the detuning frequency $\Delta \protect%
\nu {\rm ,}$ and decay time constant $T_{\protect\varphi }=0.5~$%
%TCIMACRO{\U{b5}}%
%BeginExpansion
$\mu$%
%EndExpansion
{\rm s}.}
\end{figure}

%\FRAME{dhFUw}{3.25in}{2.5391in}{0pt}{\Qcb{Figure 6: }}{}{u:/hp/papers/quantronium/ramsey.eps}{\special{language
%"Scientific Word";type "GRAPHIC";maintain-aspect-ratio TRUE;display
%"USEDEF";valid_file "F";width 3.25in;height 2.5391in;depth
%0pt;original-width 3.0701in;original-height 2.3981in;cropleft "0";croptop
%"1";cropright "1";cropbottom "0";filename 'ramsey.eps';file-properties
%"XNPEU";}}

\begin{figure}
\centerline{\epsfxsize=6.2509in \epsfysize=4.5558in
\epsffile{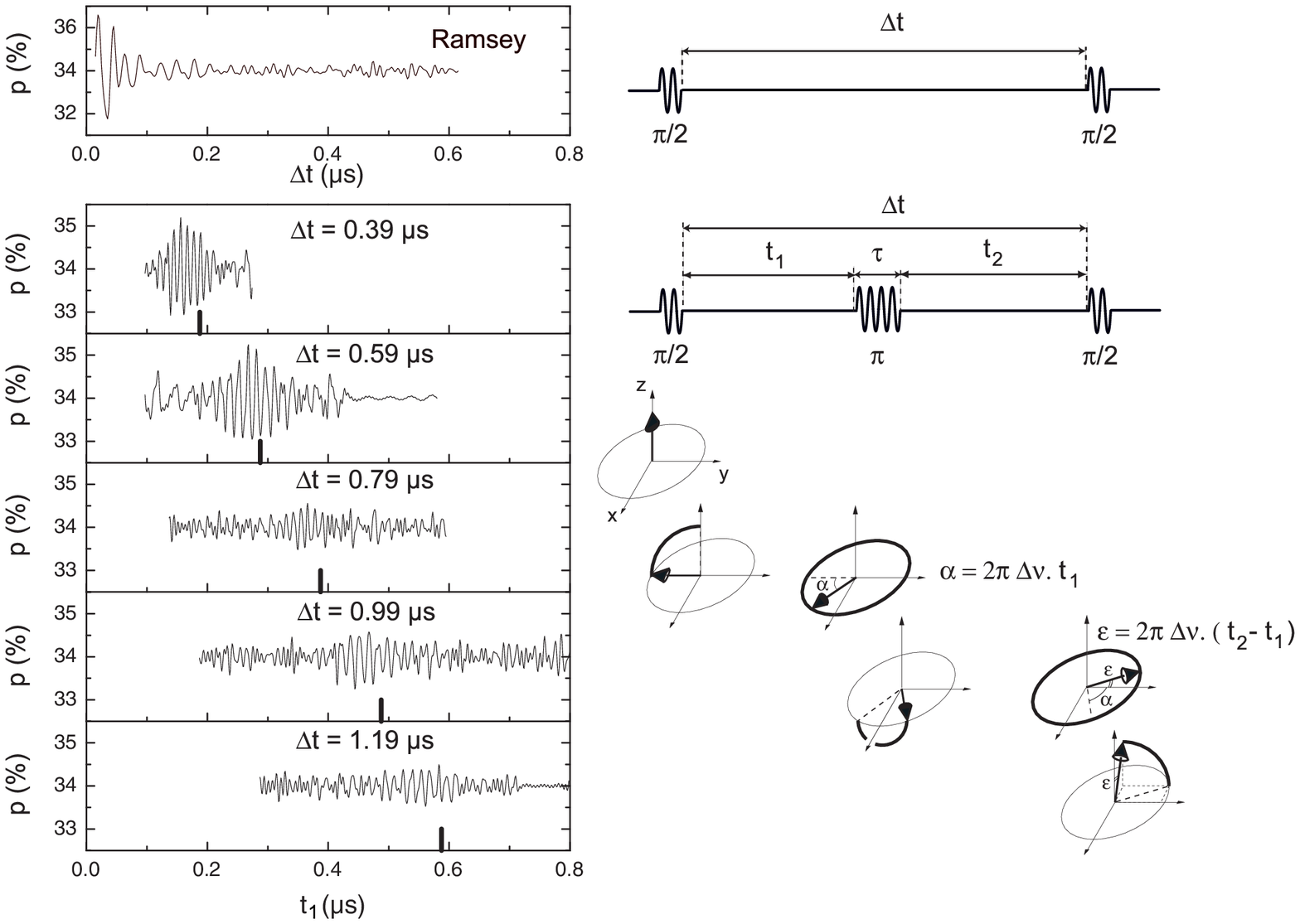} }
\caption{Top panel: Ramsey
fringes measured at $N_{g}=0.52,$ $\protect\phi =0$ and $\Delta \protect\nu %
=41~{\rm MHz.}$ The decay time constant of the fringes is here $T_{\protect%
\varphi }\sim $ 30~ns. Lower panels~: echo signals obtained with the pulse
sequence schematically described on the right side, for various sequence
durations $\Delta t$. A first $\protect\pi /2$ pulse brings the spin $\vec{s}
$ on the $-y$ axis. Follows a free precession by an angle $\protect\alpha =2%
\protect\pi \Delta \protect\nu \ t_{1}$ during a time $t_{1}$. A subsequent $%
\protect\pi $ pulse brings the spin in the symmetric position with respect
to $x$ axis. Follows a second free precession during time $t_{2}$, which
brings the spin at an angle $\protect\varepsilon =2\protect\pi \Delta 
\protect\nu (t_{2}-t_{1})$ with the $y$ axis. The last $\protect\pi /2$
pulse results in a final $z$ component of the spin equal to $\cos \protect%
\varepsilon .$ The average switching probability $p=(1-\left\langle \cos 
\protect\varepsilon \right\rangle )/2$, obtained by repeating the sequence,
is an oscillating function of $t_{2}-t_{1}.$ The amplitude of the
oscillations is damped away from $t_{1}=t_{2}$ (thick tick in each panel)
due to fluctuations of $\Delta \protect\nu .$}
\end{figure}

%\FRAME{dhFUw}{6.2509in}{4.5558in}{0pt}{\Qcb{Figure 7: }}{}{echo.eps}{\special%
%{language "Scientific Word";type "GRAPHIC";maintain-aspect-ratio
%TRUE;display "USEDEF";valid_file "F";width 6.2509in;height 4.5558in;depth
%0pt;original-width 7.9597in;original-height 11.8618in;cropleft "0";croptop
%"1";cropright "1";cropbottom "0";filename 'echo.eps';file-properties
%"XNPEU";}}

When the circuit is biased away from the optimal point, the
coherence time $T_{\varphi }\,$ of the oscillation is strongly reduced, as
shown in the top panel of Fig.~7 for $N_{g}=1/2+0.02$, $\phi =0$. In order
to determine the contribution to dephasing of low frequency charge noise, we
have performed spin echo experiments: an intermediate $\pi $ pulse is
inserted between the two $\pi /2$ pulses of the Ramsey sequence (see Fig.~7,
right side). The effect of the $\pi $ pulse is to make the phases
accumulated during the two free evolution time intervals $t_{1}$ and $t_{2}$
to subtract one from the other (see Fig.~7). By symmetry, when $t_{1}=t_{2},$
the total phase accumulated is independent of $\Delta \nu $ if $\Delta \nu $
is constant over the complete sequence. Compared to the Ramsey fringes
experiment, where coherence during $\Delta t$ is revealed by the periodic
evolution of $\left\langle \cos \left[ 2\pi \Delta \nu \Delta t\right]
\right\rangle ,$ the echo signal varies as $(1-\left\langle \cos \left[ 2\pi
\Delta \nu (t_{2}-t_{1})\right] \right\rangle )/2$ and is therefore less
sensitive to fluctuations of $\Delta \nu $ from sequence to sequence when $%
t_{1}\sim t_{2}$. In the experiment, we have recorded the switching
probability at fixed values of $\Delta t,$ as a function of the delay $t_{1}$
(left panels of Fig.~7). Up to $\Delta t\simeq 1~$ 
%TCIMACRO{\U{b5}}%
%BeginExpansion
$\mu$%
%EndExpansion
s, fringes emerge around $t_{1}=t_{2}=(\Delta t-\tau )/2$ (here, $\tau \sim
15~{\rm ns}$), indicating that during pulse sequences of this duration,
coherence was at least partly conserved. As expected, the period of the
oscillations is twice as short in the echo experiment than in the Ramsey
experiment. The observation of spin echoes at time scales much larger than
the decay time of the Ramsey fringes indicates that in this situation
decoherence was essentially due to charge fluctuations at frequencies lower
than $1/\Delta t\approx 1~{\rm MHz}${\em . }No echo was seen in experiments
performed at $\phi \neq 0$, suggesting that the relevant phase noise was at
higher frequencies.

\bigskip In all our time domain experiments, the oscillation period of the
switching probability closely agrees with theory, meaning a precise control
of the preparation of $\vec{s}$ and of its evolution. However, the amplitude
of the oscillations is smaller than expected by a factor of three to four.
This loss of contrast is likely to be due to a relaxation of the level
population during the measurement itself. In principle the current pulse,
whose rise time is 50~ns, is sufficiently adiabatic not to induce
transitions directly between the two levels. Nevertheless, it is possible
that the readout or even the preparation pulses excite resonances in the
bias circuitry which in turn could induce transitions in our two-level
manifold. Experiments using better shaped readout pulses and a bias
circuitry with better controlled high-frequency impedance are needed to
clarify this point.

In conclusion we have designed and operated a superconducting tunnel
junction circuit which behaves as a tunable two-level atom that can be
decoupled from its environment. When the readout is off, the coherence of
this ``quantronium" atom is of sufficient quality ($Q_{\varphi }=2.5\times
10^{4}$) that an arbitrary quantum evolution can be programmed with a series
of microwaves pulses. Coupling several of these circuits can be achieved
using on-chip capacitors. The ability to tune and address them individually
would allow to produce entangled states and probe their quantum
correlations. These fundamental physics experiments could lead to the
realization of quantum logic gates, an important step towards the practical
implementation of solid-state quantum processors \cite{QC}.

\bigskip \bigskip

Acknowledgements: The indispensable technical work of Pief Orfila is
gratefully acknowledged. This work has greatly benefited from direct inputs
from J. M.\ Martinis and Y. Nakamura. The authors acknowledge discussions
with P. Delsing, G. Falci, D. Haviland, H.\ Mooij, R.\ Schoelkopf, G.\ Sch%
\"{o}n and G. Wendin. This work is partly supported by the European Union
through contract IST-10673 SQUBIT and by the Conseil G\'{e}n\'{e}ral de
l'Essonne (EQUM project).


\begin{thebibliography}{99}
\bibitem{noteangle} In practice, the rotation axis does not need to be $x$,
but the rotation angle of the two pulses is always adjusted so as to bring a
spin initially along $z$ into a plane perpendicular to $z$.
\end{thebibliography}
\end{document}